\documentclass[prl,twocolumn]{revtex4-1}

\usepackage{graphicx}
\usepackage{epstopdf} 
\DeclareGraphicsExtensions{.eps, .pdf}
\usepackage{amssymb, amsmath}

\begin{document}

\newcommand{\bra}[1]{\ensuremath{\langle#1 |}}
\newcommand{\ket}[1]{\ensuremath{ |#1\rangle}}
\newcommand{\bracket}[2]{\ensuremath{\langle#1 \vphantom{#2} |  #2 \vphantom{#1} \rangle}}

\def\O{{\cal{O}}}
\def\t{\theta}
\def\H{\tilde{H}}
\def\E{\tilde{E}}
\def\P2{\tilde{P}^2}
\def\X2{\tilde{X}^2}
\def\tphi{\tilde{\phi}}
\def\sinc{\text{sinc}}
\def\rect{\text{rect}}
\def\dH{\tilde{\cal{H}}}
\def\hphi{\hat{\phi}}

\title{Electron-Phonon Systems on a Universal Quantum Computer}
\author{Alexandru Macridin, Panagiotis Spentzouris, James Amundson, Roni Harnik }
\affiliation{Fermilab, P.O. Box 500, Batavia, Illinois 60510, USA}

\begin{abstract}

We present an algorithm that extends existing quantum algorithms for simulating fermion systems in quantum chemistry and condensed matter physics 
to include bosons in general and phonons in particular. We introduce a qubit representation for the  low-energy subspace of phonons which 
allows an efficient simulation of the evolution operator of the
electron-phonon systems. As a consequence of the Nyquist-Shannon sampling theorem, the phonons are represented with exponential accuracy on a discretized  Hilbert space 
with a size that increases linearly with the cutoff of the maximum phonon number.
The additional number of qubits required by the 
presence of phonons scales linearly with the size of the system.
The additional circuit depth is constant for systems with finite-range electron-phonon and phonon-phonon interactions
and linear  for long-range electron-phonon interactions. 
Our algorithm for a Holstein polaron problem was implemented on an Atos Quantum Learning Machine (QLM) quantum simulator employing the Quantum Phase
Estimation method. The energy and the phonon number distribution of the polaron state agree  with exact diagonalization 
results for weak, intermediate and strong electron-phonon coupling regimes.

\end{abstract}

\maketitle

\paragraph{\bf Introduction.}

The algorithms for simulating many-fermion systems on quantum 
computers have progressed tremendously in recent years~\cite{abrahams_lloyd_prl_1997,abrahams_lloyd_prl_1999, ortiz_pra_2001,somma_gubernatis_2002,somma_qic_2003,
whitfield_2011,troyer_pra_2015,peruzzo_nature_2014,mcclean_NJPhys_2016}.
Due to the relatively small amount  of resources required,  near-future quantum simulations of strongly-correlated electrons
are  expected to have significant scientific impact  in quantum chemistry and condensed matter physics.
In this letter and in Ref.~\cite{macridin_fb_2018} we extend the existing fermion algorithms to 
include bosons, opening up the possibility for quantum simulation
to whole new classes of physical systems.

The electron-phonon model is  an example of non-relativistic quantum field theory.
The phonons are the most common bosonic excitations  in solids.
Their interaction with electrons  can significantly renormalize
the electric and transport properties of materials or
can lead to dramatic effects, such  as superconductivity or Jahn-Teller distortions. 
Moreover,  the interaction  of electrons with other  bosonic collective excitations
in solids (such as spin, orbital, charge, etc.) can be addressed  by similar Hamiltonians.

The quantum computation of fermion-boson systems has previously been addressed  
in trapped ion systems~\cite{trapped_ions1,trapped_ions2,trapped_ions3,trapped_ions4},
where the boson space was mapped on the ions' vibrational space. 
Our approach 
to quantum computation of systems with bosons 
is different, since  we consider
boson representation on qubits.  
While there are established ways to map fermion states to qubits~\cite{ortiz_pra_2001, whitfield_2011,kitaev_bravy},
much less is discussed about bosons. In Ref.~\cite{Lidar} bosons are
represented as a sum  of $n_x$  parafermions (qubits),  up to an error  $\O(n/n_x)$, where $n$ is the boson state occupation number.  
This  requires a large number 
of qubits, especially in the intermediate and strong coupling regimes where $n$ is large. 
In Refs.~\cite{somma_qic_2003,batista_ortiz_2007} systems with a fixed number of bosons are addressed,
but the method is not suitable to fermion-boson interacting systems where the number of bosons is not conserved.
An  algorithm for calculating scattering amplitudes  in  quantum field theories,
based on the discretization of the continuous field value at each lattice site 
has been  proposed in Ref.~\cite{jordan_science_2012}. In their approach 
the required number of qubits scales as  $\log(1/\epsilon)$, 
whereas in our   scales exponentially faster, $\approx \log(\log(1/\epsilon))$, 
where $\epsilon$ is the desired accuracy. 
We find that only a small number of 
additional qubits per site,  $n_x \approx 6\rm{~or~}7$, is enough  to simulate weak, intermediate, and strong coupling regimes
of most electron-phonon problems of interest.

We  treat the phonons as a finite set of harmonic oscillators (HO). 
We show that the low-energy space of a HO
is, up to an exponentially small error,  
isomorphic  with the low-energy subspace of a finite-sized Hilbert space. 
Similar finite-sized Hilbert space truncation is  employed by the Fourier grid Hamiltonian 
(FGH) method~\cite{ marston_kurti_jcphys_1989} and is related to more general discrete variable representation  
(DVR) methods~\cite{light_jcphys_1985, Littlejohn_2002,bulgac_forbes_prc_2013}. 
We present a novel explanation for the exponential accuracy of the FGH method 
based on the Nyquist-Shannon (NS) sampling theorem~\cite{nyquist-shanon}.
The finite-sized phonon Hilbert space is mapped onto the qubit space  of universal quantum computers.
The size of the low-energy subspace is given by the maximum phonon number cutoff; 
the size of the truncated space increases  linearly with this cutoff.
The number of qubits necessary to store phonons scales logarithmically with the cutoff
and linearly with the system size $N$.
The electrons are mapped to qubit states  via the Jordan-Wigner transformation~\cite{jordan_wigner_1928,ortiz_pra_2001, whitfield_2011}.
The algorithm simulates the evolution operator of the electron-phonon Hamiltonian.
For long-range  interactions, the additional circuit depth and the number of gates due to the 
inclusion of phonons is at worst $\O(N^2)$, while 
for finite-range  interactions the additional circuit depth is constant.

We benchmark our algorithm by running a simulation of the two-site  Holstein  polaron~\cite{holstein_1959} utilizing
the Quantum Phase Estimation (QPE) method~\cite{kitaev_qpe,cleve_1998, abrahams_lloyd_prl_1999, kitaev_qpe_2002, nielsen_2010, guzik_science_2005} 
on an Atos Quantum Learning Machine (QLM) simulator. The energy and phonon distribution of the polaron state
agree with results obtained from exact diagonalization.

%
%

\paragraph{\bf The electron-phonon model.}
The Hamiltonian is
\begin{equation}
\label{eq:ham}
H=H_e+H_{p}+H_{ep},
\end{equation}
\noindent with
\begin{equation}
\label{eq:hame}
H_e= \sum_{ij} t_{ij} \left( c^{\dagger}_i c_j +c^{\dagger}_j c_i  \right) + \sum_{ijkl} U_{ijkl}  c^{\dagger}_i c^{\dagger}_j c_k c_l,
\end{equation}
\begin{equation}
\label{eq:hamp}
H_p=\sum_{n\nu} \frac{P^2_{n\nu}}{2 M_{\nu} } + \frac{1}{2} M_{\nu}\omega^2_{n\nu} X^2_{n\nu} + \sum_{n\nu m \mu}K_{n\nu m\mu} X_{n\nu} X_{m\mu},
\end{equation}
\begin{equation}
\label{eq:hamep}
H_{ep}= \sum_{ij n \nu} g_{ijn\nu} \left( c^{\dagger}_i c_j +c^{\dagger}_j c_i  \right) X_{n \nu},
\end{equation}
\noindent where $H_e$ ($H_p$) contains electronic (phononic) degrees of freedom and  $H_{ep}$ 
describes the electron-phonon interaction. The sums are taken over the electron orbitals ($i$, $j$, $k$, $l$),  ion positions ($m$, $n$)
and vibrational modes ($\mu$, $\nu$).

\paragraph{\bf Phonon space truncation.}

The phonons in Eq.(\ref{eq:ham}) are described by a set of HOs. 
The phonon Hilbert space is a direct product of HO  spaces.
Below we address the truncation of the HO space on a finite-sized space.

\begin{figure}
\begin{center}
\includegraphics*[width=3.4in]{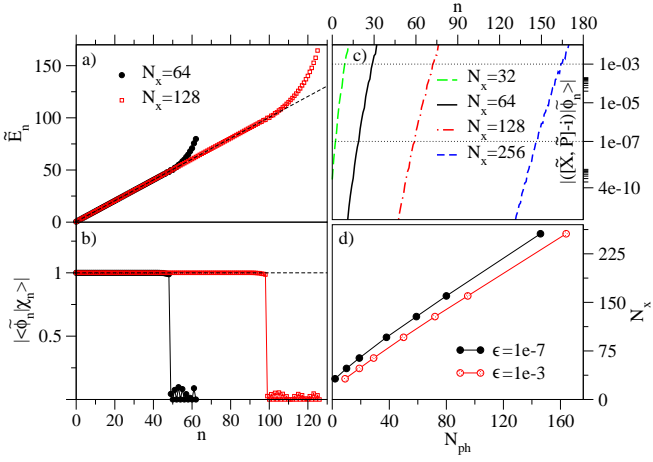}
\caption{
(a)  Eigenspectrum $\tilde{E}_n$ of  $\H_h$ (\ref{eq:tilde_ham}) for $N_x=64$ 
and $N_x=128$.
(b) Overlap between the eigenvectors  $\ket{\tphi_n}$ of $\H_h$,   
and  $\ket{\chi_n}$ (Eq.(\ref{eq:dhg})).
(c) $| ([\tilde{X}, \tilde{P} ] -i ) \ket{\tphi_n}|$ versus $n$ for different values of $N_x$.
For $n<N_{ph}$ where $N_{ph}$ is a cutoff number increasing with increasing $N_x$,
$\tilde{E}_n = n+\frac{1}{2} +\epsilon$, $\ket{\tphi_n} = \ket{\chi_n}+\epsilon$ and 
$[\tilde{X}, \tilde{P} ]\ket{\tphi_n} = i\ket{\tphi_n}+\epsilon$, 
with $\epsilon$ given by Eq.~(\ref{eq:epsilon}).
(d) The size of the discrete space, $N_x$, increases linearly with the size of the low-energy subspace, $N_{ph}$.
The full (open) symbols are extracted from (c) for $\epsilon=10^{-7}$ ($\epsilon=10^{-3}$).}
\label{fig:trunc}
\end{center}
\end{figure}

The HO Hamiltonian is $H_h=P^2/2+X^2/2$,
where the operators $X$, $P$ and $H_h$ are rescaled by $1/\sqrt{M \omega}$, $\sqrt{M \omega}$ and
 $1/\omega$, respectively. The   eigenspectrum and the eigenvectors in the position basis  are
\begin{equation}
\label{eq:hos_en}
E_n = n+\frac{1}{2}, ~ \bracket{x}{\phi_n} \equiv \phi_n(x) = \frac{1}{\pi^{\frac{1}{4}}\sqrt{2^n n!}} e^{-\frac{x^2}{2}}H_n(x).
\end{equation}
\noindent The Hermite-Gauss (HG) functions $\phi_n(x)$
are also eigenfunctions of the Fourier transform operator~\cite{FT_HG}, 
\begin{equation}
\label{eq:fthg}
[{\cal{F}} (\phi_n)](p) \equiv \hat{\phi}_n(p)=(-i)^n\phi_n(p).
\end{equation}
\noindent and satisfy 
\begin{eqnarray}
\label{eq:xhg}
x\phi_n(x)=\left( \sqrt{n+1} \phi_{n+1}(x) + \sqrt{n} \phi_{n-1}(x) \right)/\sqrt{2} \\
\label{eq:phg}
p \hat{\phi}_n(p)=i\left( \sqrt{n+1}   \hat{\phi}_{n+1}(p) - \sqrt{n} \hat{\phi}_{n-1}(p) \right)/\sqrt{2}.
\end{eqnarray}
\noindent The equations (\ref{eq:xhg}) and (\ref{eq:phg}) are the  eigenvalue equations 
for the position $X=\left( b^{\dagger}+ b \right)/\sqrt{2}$
and momentum $P=i\left( b^{\dagger} - b \right)/\sqrt{2}$ operators, 
where  $b^{\dagger}$  ($b$) is the creation (annihilation) operator.

The HG functions  fall exponentially fast to zero for large argument. 
For any positive integer cutoff $N_{ph}$, 
a half-width $L$ can be chosen such that  for all $n<N_{ph}$,
$|\hphi_n(p)| < \epsilon$ for  $|p|>L$ and  $|\phi_n(x)| < \epsilon$ for  $|x|>L$ ,
where $\epsilon \propto \exp (-L^2/2 )$.  
With exponentially good accuracy we can restrict to the region $|p|<L$ and $|x|<L$.
The NS sampling theorem~\cite{nyquist-shanon} states that, without  loss of information,  $\phi_n(x)$ can be sampled at 
points $x_i=i \Delta$, where $i$ is an integer
and $\Delta=\pi/L$.
We can restrict $i$  to $N_x$ sampling points, $i=\overline{-N_x/2, N_x/2-1}$, such that $|x|<L$.
This implies $2L = N_x \Delta=\sqrt{2 \pi N_x}$~\cite{macridin_fb_2018}.

Let us consider the $N_x$ finite-sized subspace, $\tilde{\cal{H}}$, spanned by the sampling  position vectors 
$\{\ket{x_i}\}_i$, and define the vectors $\ket{\chi_n} \in \dH$ by
\begin{eqnarray}
\label{eq:dhg}
\bracket{x_i}{\chi_n} \equiv \sqrt{\Delta} \phi_n(x_i).
\end{eqnarray}
\noindent As a consequence of the  NS theorem~\cite{macridin_fb_2018}, the vectors
$\{ \ket{\chi_n}\}_{n<N_{ph}}$ are orthonormal and
\begin{eqnarray}
\label{eq:pdhg}
\bracket{p_m}{\chi_n}=\sqrt{2 \pi \Delta}\hat{\phi}_n(p_m),
\end{eqnarray}
\noindent where $\ket{p_m} = N^{-1/2}_x\sum_{i=-\frac{N_x}{2}}^{ \frac{N_x}{2}-1 } e^{i x_i p_m} \ket{x_i}$.
In Eq.(\ref{eq:pdhg}) $\hat{\phi}_n(p_m)$ is the HG function in the momentum representation (Eq.(\ref{eq:fthg})) sampled at 
$p_m=m \Delta$ with $m=\overline{-N_x/2, N_x/2-1}$.

Since  $\bracket{x_i}{\chi_n} \propto \phi_n(x_i)$
and $\bracket{p_m}{\chi_n} \propto \hat{\phi}_n(p_m)$,  Eqs.(\ref{eq:xhg}), (\ref{eq:phg}), (\ref{eq:dhg}) and (\ref{eq:pdhg}) imply
\begin{equation}
\label{eq:xchi}
x_i \bracket{x_i}{\chi_n}=\left( \sqrt{n+1} \bracket{x_i}{\chi_{n+1}} + \sqrt{n} \bracket{x_i}{\chi_{n-1}} \right)/\sqrt{2},
\end{equation}
\begin{equation}
\label{eq:pchi}
p_m \bracket{p_m}{\chi_n}= i\left( \sqrt{n+1} \bracket{p_m}{\chi_{n+1}} - \sqrt{n} \bracket{p_m}{\chi_{n-1}} \right)/\sqrt{2},
\end{equation}
\noindent for $n<N_{ph}$.
If we define the  operators
\begin{eqnarray}
\label{eq:tilde_x}
\tilde{X} \ket{x_i} =x_i \ket{x_i}, \\
\label{eq:tilde_p}
\tilde{P} \ket{p_m} =p_m \ket{p_m},
\end{eqnarray}
\noindent acting on $\dH$,  Eqs.(\ref{eq:xchi}) and (\ref{eq:pchi}) read
\begin{eqnarray}
\label{eq:txchi}
\tilde{X} \ket{\chi_n}=\left( \sqrt{n+1} \ket{\chi_{n+1}} + \sqrt{n} \ket{\chi_{n-1}} \right)/\sqrt{2},\\
\label{eq:tpchi}
\tilde{P} \ket{\chi_n}=i\left( \sqrt{n+1} \ket{\chi_{n+1}} - \sqrt{n} \ket{\chi_{n-1}} \right)\sqrt{2},
\end{eqnarray}
\noindent which implies $[\tilde{X}, \tilde{P}] \ket{\chi_n}=i \ket{\chi_n}$ for $n<N_{ph}$.
On the  subspace  spanned by $\{ \ket{\chi_n}\}_{n<N_{ph}}$ one has $[\tilde{X}, \tilde{P}]=i$.
Therefore the algebra generated by  $\tilde{X}$ and  $\tilde{P}$ is isomorphic with the algebra generated by $X$ and $P$
on the harmonic oscillator subspace spanned by  $\{ \ket{\phi_n}\}_{n<N_{ph}}$. 

The vectors $\{ \ket{\chi_n}\}_{n<N_{ph}}$  are eigenvectors of 
\begin{eqnarray}
\label{eq:tilde_ham}
\H_h=\P2/2+ \X2/2,
\end{eqnarray}
\noindent satisfying  $\H_h\ket{\chi_n}= \left(n+1/2\right)\ket{\chi_n}$. 
Moreover,  they span the low-energy subspace of $\dH$,  
as the numerical investigation presented below shows.

The eigenspectrum $\tilde{E}_n$ of $\H_h$ calculated by exact diagonalization
is shown in Fig.~\ref{fig:trunc}(a). The first $N_{ph}$ energy levels are the same 
as the corresponding HO energy levels, {\em i.e.,} $\tilde{E}_n=n+1/2+\epsilon$.
The eigenstates  $\{ \ket{\tphi_n}\}_{n<N_{ph}}$ of $\H_h$ are
the projected HG functions on the discrete basis $\{\ket{\chi_{n}}\}_{n<N_{ph}}$, Eq.(\ref{eq:dhg}).
This can be inferred from Fig.~\ref{fig:trunc}(b) 
where we see that the overlap $| \bracket{\tphi_{n}}{\chi_n} | =1 - \epsilon$ for $n<N_{ph}$. 
Fig.~\ref{fig:trunc}(c) shows that $|([\tilde{X}, \tilde{P}]-i) \ket{\tphi_n} | < \epsilon$
for $n < N_{ph}$. The value of $\epsilon$ is exponentially small, a consequence of
cutting the tails of the  HG functions for $| x |, | p |>L$. Numerically, we find  
\begin{equation}
\label{eq:epsilon}
\epsilon  \lesssim 10 \exp[-(0.51 N_x -0.765 N_{ph})].
\end{equation}
The numerical results agree with the analytical predictions, 
supporting the isomorphism  between the $\{ \tilde{X}, \tilde{P} \} $ 
and the $\{X, P \}$ algebras  on the low-energy subspace defined by $n < N_{ph}$.

The size  $N_x$ of  $\dH$  increases approximately linearly with increasing $N_{ph}$.
In Fig.~\ref{fig:trunc}(d) we plot the minimum $N_x$ necessary to  have  $N_{ph}$ states
in the low-energy regime with $\epsilon=10^{-7}$  and $\epsilon=10^{-3}$ accuracy. 
The proportionality between $N_x$ and  $N_{ph}$ is a consequence 
of the relations $L_{N_{ph}} \underset{\sim}{\propto}  \sqrt{N_{ph}}$~\cite{macridin_fb_2018} 
and $L_{N_{ph}} \propto  \sqrt{N_{x}}$.

As long as the physics can be addressed by truncating the number of phonons
per state our finite-sized representation  is suitable for computation. 
The cutoff $N_{ph}$ increases with increasing effective strength of interaction.  
For stable systems the truncation errors are expected to converge exponentially
quickly to zero with increasing $N_{ph}$~\cite{macridin_fb_2018}.

\paragraph{\bf Algorithm.}

Our algorithm  simulates  the evolution operator $\exp(-i H t)$ on a gate quantum computer.
We employ the Trotter-Suzuki expansion~\cite{trotter_1959, suzuki_1976} of $\exp(-i H t)$
to a product of  short-time evolution operators corresponding to the
noncommuting terms in the Hamiltonian.

On a gate quantum computer each HO state is represented
as a superposition of  $N_x$ discrete states $\{\ket{x}\}$ 
and stored on a register of $n_x=\log_2{N_x}$ qubits. 
The operators $X$ and $P$  are 
replaced by their discrete 
versions $\tilde{X}$ (Eq.(\ref{eq:tilde_x})) and $\tilde{P}$ (Eq.(\ref{eq:tilde_p})), respectively. 
The following equations are true: $\tilde{X}\ket{x}=\tilde{x}\ket{x}$ and $\tilde{P}\ket{p}=\tilde{p}\ket{p}$, where
$\{\ket{p}\}$ are obtained from $\{\ket{x}\}$ via the discrete Fourier transform.
The eigenvalues are $\tilde{x}=(x-N_x/2) \Delta$ and $\tilde{p}=[(p+N_x/2) \mod N_x-N_x/2] \Delta$. They are different from the ones in  
Eqs.~(\ref{eq:tilde_x}) and (\ref{eq:tilde_p}) since the stored states in the qubit registers
are numbers  between $0$ and $N_x-1$ and not between $-N_x/2$ and $N_x/2-1$. 

\paragraph{Phonon evolution.}

\begin{figure}[tb]
\begin{center}
\includegraphics*[width=3.4in]{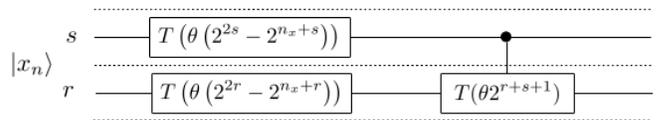}
\caption{The circuit 
$\ket{x_n} \longrightarrow \exp(i 2^{n_x-2} \t) \exp[-i (x_n-2^{n_x-1} )^2\t] \ket{x_n}$
requires $n_x$ phase shift gates  and $n_x (n_x-1)/2$  controlled  phase shift gates.
The angles of the phase shift gates are determined by writing 
$(x_n-2^{n_x-1})^2=\sum^{n_x-1}_{r=0} x_n^r \left( 2^{2r}-2^{n_x+r} \right) +\sum_{r<s} x_n^r x_n^s 2^{r+s+1}+2^{2 n_x-2}$, 
where $\{x_n^r\}_{r=\overline{0,n_x-1}}$ is the binary representation of $x_n$.}
\label{fig:ps_x2}
\end{center}
\end{figure}

Within the Trotter approximation, the algorithm for the evolution of
phonons requires the implementation of $\exp(-i \t \X2_n) \ket{x_n}$,  $\exp(-i \t \P2_n)\ket{x_n}$ and 
$\exp(-i \t \tilde{X_n} \tilde{X_m})\ket{x_n}\ket{x_m}$, where $n$ and $m$ are HO labels.

The implementation of $\exp(-i \t \X2_n) \ket{x_n}$ 
requires phase shift gates $T$ and is shown in Fig.~\ref{fig:ps_x2}.
The angles of the phase shift gates are 
determined by  writing the eigenvalues of $\X2_n$ in 
binary format, as shown in the figure's caption.
A phase factor equal to $\exp(i 2^{n_x-2} \t)$ 
accumulates at each Trotter step.  This phase factor can be tracked classically.

For the implementation of $\exp(-i \t \P2_n)\ket{x_n}$ one first applies a  quantum Fourier transform (QFT)~\cite{nielsen_2010}
$\ket{x_n}\xrightarrow{QFT}  \ket{p_n}$, an idea  first discussed in Refs.~\cite{zalka_1998,wiesner_1996}. 
Then  $\exp(-i \t \P2_n)\ket{p_n}$ 
is implemented by a circuit similar to the one shown in Fig~\ref{fig:ps_x2}. The last step is
an inverse QFT, $\ket{p_n}\xrightarrow{IQFT}  \ket{x_n}$.

The  operator $\exp(-i \t \tilde{X}_n \tilde{X}_m)\ket{x_n}\ket{x_m}$
requires two phonon registers, $n$ and $m$.
The phase shift angles are determined by writing the product $\tilde{x}_n \tilde{x}_m$ as a sum with binary coefficients~\cite{macridin_fb_2018}. 
The circuit is similar to the one in Fig.~\ref{fig:ps_x2}. It has $n^2_x$  controlled phase shift gates and $2 n_x$  phase shift gates.

\begin{figure}[tb]
\begin{center}
\includegraphics*[width=3.4in]{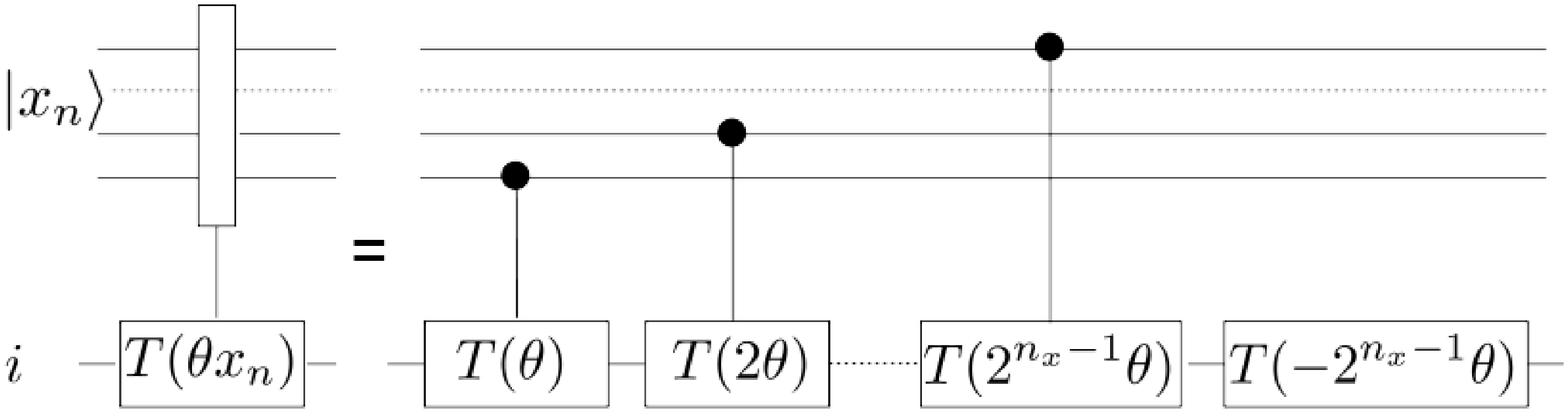}
\caption{ Circuit for $\exp(-i \t c^{\dagger}_i c_i \tilde{X}_n) \ket{i}  \otimes \ket{x_n}$.
The phase shift angle is
 $\t(x_n-N_x/2)=\t\sum^{n_x-1}_{r=0} x^r_n 2^r-\t 2^{n_x-1}$, where $\{x^r_n\}_{r=\overline{0,n_x-1}}$ take binary values.}
\label{fig:pase_sf_x}
\end{center}
\end{figure}

\paragraph{Electron evolution.} 
The algorithm for fermions is described at length in numerous papers (see Refs.~\cite{somma_gubernatis_2002, whitfield_2011,troyer_pra_2015}.)
We assume  here a Jordan-Wigner mapping of the fermion operators to
the Pauli operators $X$, $Y$, and $Z$ as in Ref.~\cite{troyer_pra_2015}.
Each electron orbital requires a qubit, the state $\ket{\uparrow} \equiv \ket{0}$ ($\ket{\downarrow} \equiv \ket{1}$) corresponding
to an unoccupied (occupied) orbital.

\paragraph{Interaction term evolution.} 
The implementation of the electron-phonon interaction  is similar to the one for 
single-particle electron  operators which 
requires phase shift $T(\t)$ or z-rotations $R_z(\t)$ gates acting on the electron qubits~\cite{whitfield_2011,troyer_pra_2015}.
The difference is
the value of the gate angle $\t$,  which  is replaced by $\t \tilde{x}$, where $\tilde{x}$ is the eigenvalue of $\tilde{X}$ corresponding to the phonon state $\ket{x}$.

In  Fig.~\ref{fig:pase_sf_x} we show the implementation of 
$\exp(-i \t c^{\dagger}_i c_i \tilde{X}_n) \ket{i} \otimes \ket{x_n}=\left( T(\theta \tilde{x}_n) \ket{i} \right) \otimes \ket{x_n}$ 
\noindent where $\ket{i}$ is the  $i$ fermion orbital  and $\ket{x_n}$ is the state of the HO $n$.

The circuit for  $\exp(-i \t \left( c^{\dagger}_i c_j+ c^{\dagger}_j c_i \right) \tilde{X}_n)$ (not shown)
is similar to the circuit shown in Fig.~(9) of Ref.~\cite{troyer_pra_2015} or Table~A1 of Ref.~\cite{whitfield_2011}
for $\exp[-i \t ( c^{\dagger}_i c_j+ c^{\dagger}_j c_i ) ]$.  The difference is that  $R_z(\t)$ is replaced by  $R_z(\t \tilde{x}_n)$ 
(see Fig.~8 in Ref.~\cite{macridin_fb_2018}).

The nonlocality of the Jordan-Wigner mapping 
increases the circuit depth for fermion algorithms~\cite{somma_gubernatis_2002, whitfield_2011, troyer_pra_2015}.
However, the implementation of the  electron hopping and electron-phonon terms
can be combined. One can implement
$\exp[-i(  c^{\dagger}_i c_j + c^{\dagger}_i c_j  ) (\t_0 +\sum_n\t_n \tilde{X}_n ) ]$,
and there will be no  additional Jordan-Wigner strings due to electron-phonon terms.
The contribution to the circuit depth for long-range electron-phonon interactions is $\O(N)$.

\paragraph{Input state preparation.}
The input state for the QPE algorithms must have 
a large overlap with the ground state. The input state can be obtained by
the adiabatic method~\cite{Farhi_science_2001}, starting with 
$H_0=H_e+H_{p}$
and slowly turning on the electron-phonon interaction.
The ground state of $H_0$ is $\ket{f_0}  \otimes \ket{\Phi_0}$, where  
$\ket{f_0} $ is
the fermion Hamiltonian ground state. Its preparation, while non-trivial,  is addressed in 
the literature~\cite{ortiz_pra_2001,lidar_prl_2002, whitfield_2011,troyer_pra_2015}.
The ground state of $H_p$
is a direct product of grid-projected  Gaussian functions $\ket{\chi_0}$, Eq.(\ref{eq:dhg}).

Methods to prepare Gaussian states are discussed in Refs.~\cite{grover_wave,kitaev_gaussian}.
However, for the polaron simulations we use the variational method
to prepare $\ket{\chi_0}$~\cite{macridin_fb_2018}. This method 
 is especially useful for near-term computation since it 
requires low-depth circuits. We find that Gaussian 
states on $n_x=6,7$ qubit registers 
can be obtained with high fidelity ($>0.998$) under the action of a $N_S=6$ step
unitary operator
\begin{eqnarray}
\label{eq:vinput}
\ket{\phi_v}= \prod_{s=1}^{N_S} U^s( \pmb{\theta}^s, \pmb{\rho}^s) \ket{x=0}.
\end{eqnarray}
\noindent The operator $U^s( \pmb{\theta}^s, \pmb{\rho}^s)$ is a product of 
$\exp(-i\rho^s_p \P2)$,  $\exp(-i\rho^s_x \X2)$ and single qubit rotations, $\exp(-i \t_{x}^s X)$, $\exp(-i \t_{y}^s Y)$ and $\exp(-i \t_{z}^s Z)$. 
The variational parameters ${\pmb{\theta}}^s=\left\{\t^s_{xi}, \t^s_{yi}, \t^s_{zi} \right\}_{i=\overline{0,n_x-1}}$ and
${\pmb{\rho}}^s=\left\{\rho^s_{x}, \rho^s_{p}\right\}$ are optimized for maximum fidelity
$|\bracket{\phi_v}{\chi_0}|^2$.

\paragraph{Measurements.}
Measurements methods described previously~\cite{somma_gubernatis_2002,troyer_pra_2015}
can be applied to our algorithm.

\paragraph{Resource scaling.} 
The number of additional qubits required by phonons is $\O(N n_x)$,
with $n_x =  \O\left(\log \left[ \ln(\epsilon^{-1})
+0.765 N_{ph}\left( \epsilon^{-1} \right)\right]\right)$ where $\epsilon$ is the target accuracy (see Eq.~(\ref{eq:epsilon}).
Since for electron-phonon systems the phonon number distribution is Poissonian,
$N_{ph}=\O(\sqrt{\ln(\epsilon^{-1})})$ (see~\cite{macridin_fb_2018}),  implying
$n_x = \O\left(\log \left[ \ln(\epsilon^{-1})\right]\right)$.
For finite-range  interactions the phonons 
introduce an $\O(N)$ contribution to the total number of gates and a constant
contribution to the circuit depth. 
For long-range electron-phonon  interactions the circuit depth increases linearly with $N$
while the additional number of gates needed is $\O(N^2)$. For long-range phonon-phonon couplings both 
the additional number of gates and the circuit depth scale as $\O(N^2)$. 

\paragraph{\bf Holstein polaron on a quantum simulator.}

\begin{figure}
\begin{center}
\includegraphics*[width=3.4in]{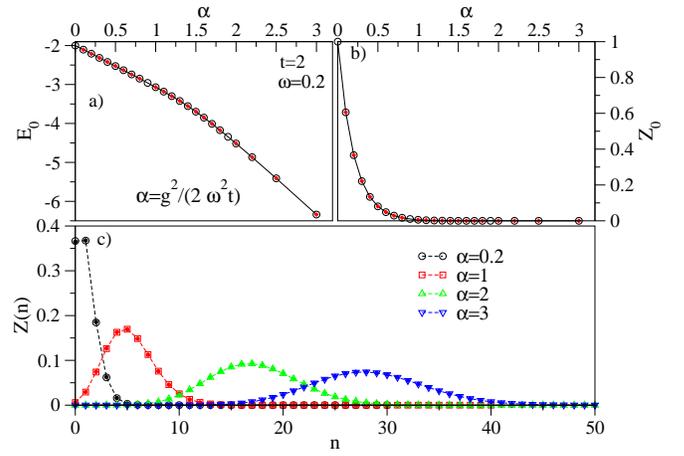}
\caption{$n_x=6$ qubits per HO. The energy (a) and quasiparticle weight (b) for the $2$-site Holstein polaron versus  coupling 
strength. 
(c) The phonon number distribution  for different  couplings. 
The open (full) symbols are computed
using exact diagonalization  (QPE algorithm on a quantum simulator).} 
\label{fig:2holstein}
\end{center}
\end{figure}

The polaron problem~\cite{landau_polaron}, {\em i.e.,} a single electron interacting with phonons, 
has been addressed extensively in the
literature. In the Holstein model~\cite{holstein_1959} the  phonons are described as  set
of independent oscillators located at every site. The  electron density  couples 
locally to the displacement of the HO,
\begin{eqnarray}
\label{eq:ev_holstein}
H= H_e + g \sum_i c^{\dagger}_i c_i X_i
+\sum_i \frac{P_i^2}{2} +\frac{1}{2} \omega^2 X_i^2.
\end{eqnarray}

To check the validity of our algorithm we ran a QPE code for the Holstein
polaron  on a $2$-site lattice using an Atos QLM simulator. 
The $2$-site polaron can be solved 
using the exact diagonalization method on a conventional computer. A comparison between
exact diagonalization and our quantum algorithm  is shown in Fig.~\ref{fig:2holstein}. 
The agreement is  good, with a difference of $\O(10^{-4})$ due mainly to the use of the
Trotter approximation. We find that $n_x=6$ qubits for each HO is 
enough to describe the physics even in the strong coupling regime, 
which in our case implies a  cutoff of $N_{ph} \approx 45$ phonons per site.

In Fig.~\ref{fig:2holstein}(a) the energy of the polaron  as a function of the
dimensionless coupling constant $\alpha=g^2/2\omega^2t$  is plotted.
Even this simple 2-site model
captures some essential features of more realistic polarons.
The transition from light to heavy polarons as a function of the 
coupling strength is smooth, similar to what is 
seen in 1D polaron models~\cite{wellein_fehske}.

The polaron state can be written as $\ket{\Phi}= \sum_{n=0} \sum_r a_{nr} \ket{n,r} $,
where $\{ \ket{n,r}\}_r$  are normalized vectors spanning the sector  with one electron and $n$ phonons.
The phonon distribution  is defined as $Z(n)=\sum_r |a_{nr}|^2$
and can be determined by applying the QPE algorithm
for the phonon evolution Hamiltonian $H_{p}=\sum_i P_i^2/2 +\omega^2 X_i^2/2$.
Since $\ket{\Phi}$ is not an eigenstate of $H_{p}$, the energy $E_n=\omega(n+1/2)$ is
measured with the probability $Z(n)$.

The quasiparticle weight $Z(0)$ as a function
of the coupling strength is  shown in Fig.~\ref{fig:2holstein} (b). This quantity represent the amount of the free electron in the polaron state
and gives the quasiparticle  weight measured in the photoemission experiments.
In Fig.~\ref{fig:2holstein} (c), $Z(n)$ is shown for several values of the coupling strength
corresponding to weak, intermediate and strong coupling regimes. 
The exact diagonalization and the QPE results agree well.

\paragraph{\bf Conclusions.}
We introduce a quantum algorithm for electron-phonon interacting systems
which  extends the existing quantum fermion algorithms to include phonons.
The phonons are represented as a set of HOs.  Each HO space
is represented on a finite-sized Hilbert space $\dH$. We define  operators $\tilde{X}$ and $\tilde{P}$
on $\dH$ and show that, in the low-energy subspace, the algebra generated by $\{\tilde{X}, \tilde{P} \}$
is, up to an exponentially small error,  isomorphic with the algebra generated by $\{X, P \}$. 
The size of the low-energy subspace  increases approximately linearly  
with increasing phonon cutoff number $N_{ph}$. 
We find that a small number of qubits, $n_x \approx 6,7$ per HO, is large enough 
for the simulation of weak, intermediate and strong coupling regimes of most electron-phonons
problems of interest.

Our algorithm maps all HO  spaces  $\dH$  on the qubit space and
simulates the evolution operator of the electron-phonon Hamiltonian. 
We present  circuits for the implementation of 
small  evolution steps corresponding to different terms in the Hamiltonian. 
The number of additional qubits required to add  phonons is $\O(N)$ where $N$ is proportional
to the system size.
For long-range  interactions, the additional circuit depth and the number of gates due to the phonon
inclusion is at worst $\O(N^2)$, while 
for finite-range  interactions the additional circuit depth is constant.

We benchmarked our algorithm on Atos QLM  simulator for a two-site Holstein  polaron.
The  polaron energy  and phonon distribution 
are in excellent agreement with the ones calculated by exact diagonalization.




\paragraph{\bf  Acknowledgments.}
We thank Andy Li, Eric Stern, Patrick Fox and Kiel Howe for discussions.
This manuscript has been authored by Fermi Research Alliance,
LLC under Contract No. DE-AC02-07CH11359 with the U.S. Department of Energy, Office of Science, Office of High Energy Physics. 
We gratefully acknowledge the computing resources provided and operated by 
the Joint Laboratory for System Evaluation (JLSE) at Argonne National Laboratory.
We would like to thank Atos for the use of their 38-Qubit Quantum Learning Machine (QLM) 
and support of their universal programming language AQASM.

\end{document}